\documentclass[a4paper]{article}

\usepackage{multirow}
\usepackage{INTERSPEECH2019}
\usepackage{enumitem}
\usepackage{footnote}
\makesavenoteenv{tabular}
\makesavenoteenv{table}
\usepackage{graphicx}
\graphicspath{{./pdf/}}
\DeclareGraphicsExtensions{.pdf}

\title{Large Margin Softmax Loss for Speaker Verification}
\name{Yi Liu, Liang He, Jia Liu}
\address{
  Tsinghua National Laboratory for Information Science and Technology, \\
  Department of Electronic Engineering, Tsinghua University, Beijing 100084, China}
\email{liu-yi15@mails.tsinghua.edu.cn, \{heliang, liuj\}@tsinghua.edu.cn}

\begin{document}

\maketitle
\begin{abstract}
  In neural network based speaker verification, speaker embedding is expected to be discriminative between speakers while the intra-speaker distance should remain small.
  A variety of loss functions have been proposed to achieve this goal.
  In this paper, we investigate the large margin softmax loss with different configurations in speaker verification.
  Ring loss and minimum hyperspherical energy criterion are introduced to further improve the performance.
  Results on VoxCeleb show that our best system outperforms the baseline approach by 15\% in EER, and by 13\%, 33\% in minDCF08 and minDCF10, respectively.
\end{abstract}

\section{Introduction}
Recently, speaker embedding, which is extracted from a deep neural network, outperforms the conventional i-vector in many speaker verification tasks \cite{xvector}.
By virtue of the excellent performance, speaker embedding is becoming the next generation of speaker recognition technology.
Similar with i-vector, speaker embedding encodes a variable-length utterance into a fix-length vector representing the speaker characteristics.
A variety of backend classifiers can be applied to suppress noise and session variability.
Speaker embedding can also be used in other applications such as speaker diarization \cite{xvector_diar}, speaker retrieval and speech synthesis.

Speaker verification is an open-set recognition problem.
An utterance is verified to be a certain speaker if their similarity exceeds a threshold.
Ideal speaker embedding should be discriminative between different speakers and compact within the same speaker.
Although cross-entropy with softmax is arguably one of the most commonly used loss function to train the speaker embedding neural network, it is more suitable for classification and does not explicitly encourage discriminative learning of features.

To address this issue, different loss functions are proposed.
Triplet loss for speaker verification were first presented in \cite{deep_speaker, triplet_speaker}.
By selecting appropriate training samples, triplet loss performed well in both text-dependent and text-independent tasks.
However, the performance is sensitive to the triplet mining strategy \cite{triplet_loss, triplet_sampling} and it is time consuming to design such a training procedure.
Also, speaker identity subspace loss \cite{subspace_loss}, Gaussian Mixture loss \cite{gmloss}, etc., were proposed in other works.

On the other hand, efforts have been made to improve the original softmax loss.
Center loss was introduced in \cite{center_loss, center_loss2} to constrain features to be gathered around the corresponding centers and thus reduce the intra-speaker variability.
Both triplet loss and center loss are optimized in the Euclidean space.
In the last few years, angular-based losses have become popular.
Compare with the Euclidean distance, angular distance is a more natural choice in the feature space.
In \cite{l2_norm, normface}, the features and the weights of the output layer were normalized before softmax, making the loss function focus on the cosine similarity.
Generalized end-to-end loss was proposed in \cite{ge2e}. The scaled cosine scores between the features and the estimated speaker centers were used as the logits to compute the loss.
Angular margin softmax loss was first presented in \cite{lmsoftmax, sphereface} in which the margin is incorporated with the angle in a multiplicative way.
This method is extended in \cite{amsoftmax, cosface, arcface} where additive margins are used.
Some of these losses have been applied in speaker verification \cite{asoftmax_speaker, asoftmax_speaker_2, amsoftmax_speaker, asoftmax_lambda, asoftmax_voxceleb}.
Since all these losses combine softmax with margins, we call them the large margin softmax loss in this paper.

In this paper, we first build a baseline system using a generic toolkit similar with \cite{but_system}.
Several training strategies are used to improve the accuracy.
We then compare the performance of the large margin softmax loss using different configurations.
Ring loss \cite{ring_loss} and minimum hyperspherical energy (MHE) criterion \cite{mhe} are involved to enhance the discriminative learning and enlarge the inter-speaker separability.
Experiments on VoxCeleb show that our baseline system achieves better performance than the Kaldi x-vector recipe and reduces the EER, minDCF08 and minDCF10 from 3.10\%, 0.0169, 0.4977 to 2.34\%, 0.0122 and 0.3754, respectively.
Using the large margin softmax loss with auxiliary objective functions, the best system further improves the performance to 2\%, 0.0106 and 0.2487.

The organization of this paper is as follows. The speaker embedding we use is briefly introduced in Section 2. Section 3 describes the large margin softmax loss and different techniques to enhance the loss. Our experimental setup and results are given in Section 4 and 5. The last section concludes the paper.

\section{Speaker embedding}
The deep neural network used to extract speaker embedding consists of frame-level and segment-level sub-networks, connected by a temporal pooling layer.
The frame-level network can be seen as a speaker feature extractor which transforms the acoustical features into speaker-related vectors.
These vectors are aggregated across the entire utterance by a pooling layer and further processed by several fully-connected layers.
Different loss functions can be used to optimize the network.
After training, the output of a hidden layer at the segment-level network is extracted as the speaker embedding.
Cosine scoring, LDA and PLDA are usually applied to generate the verification scores.
Since the output layer is removed during the test phase, the test speakers do not have to be present in the training data.

In this paper, speaker embedding is extracted from the x-vector architecture \cite{xvector}.
The x-vector is popular on many applications and has been provided as the official system on the recent NIST speaker recognition evaluation (SRE).
The details are described in Section 4.2.

\section{Large margin softmax loss}

\subsection{Definition}
The widely used softmax loss is presented as
\begin{equation}
  L_{S}= -\frac{1}{N} \sum_{i=1}^{N} \log \frac{e^{\vec{w}_{y_i}^T \vec{x}_i}}{\sum_{j=1}^{C} e^{\vec{w}_j^T \vec{x}_i}}
  \label{softmax}
\end{equation}
where $N$ is the number of training samples, $C$ is the number of speakers in the training set, $\vec{w}_j$ is the $j$-th column of the weights in the output layer. $\vec{x}_i$ is the input of the last (i.e. output) layer, $y_i$ is the ground truth label for the $i$-th sample. To avoid ambiguity, we use \textit{feature} to represent $\vec{x}_i$ in this paper while \textit{embedding} refers to the speaker embedding extracted from a hidden layer of the network.
The logit $\vec{w_j}^T \vec{x}_i$ can be transformed to $\lVert\vec{w}_j\rVert \lVert\vec{x}_i\rVert\cos{\theta_j}$, where $\theta_j$ is the angle between $\vec{w}_j$ and $\vec{x}_i$.
Eq. \ref{softmax} is influenced by the norm of the weights. This is annoying since we more care about the angle $\theta_j$.

In modified softmax \cite{sphereface}, the weights are $L_2$ normalized as $\lVert \vec{w}_j \rVert =1$ and the loss is
\begin{equation}
  L_{MS}= -\frac{1}{N} \sum_{i=1}^{N} \log \frac{e^{s \cos \theta_{y_i}}}{ \sum_{j=1}^{C} e^{s \cos \theta_j}}
  \label{modified_softmax}
\end{equation}
where $s$ is the scaling factor. This factor can be the feature norm $\lVert \vec{x}_i \rVert$ or a fixed value if the feature is also normalized. We will discuss the feature normalization later.

Based on Eq. \ref{modified_softmax}, different margins can be introduced by reformulating the target logits.
We define an angle function \cite{arcface}
\begin{equation}
  \psi(\theta_{y_i}) = \cos(m_1 \theta_{y_i} + m_2) - m_3
  \label{angle_function}
\end{equation}
where $m_1$, $m_2$ and $m_3$ are margins. Eq. \ref{modified_softmax} is rewritten as the large margin softmax loss
\begin{equation}
  L_{LMS}= -\frac{1}{N} \sum_{i=1}^{N} \log \frac{e^{s \cdot \psi(\theta_{y_i})}}{e^{s \cdot \psi(\theta_{y_i})} + \sum_{j=1,j\neq i}^{C} e^{s \cdot \cos\theta_j}}
  \label{margin_softmax}
\end{equation}
Strictly, Eq. \ref{angle_function} is only valid when $m_1=1$ and $m_2=0$, since $\forall \theta_{y_i}\in [0, \pi]$, $\psi(\theta_{y_i})$ should be a monotonically decreasing function.
However, in practice, the angle is usually in the range of $[0.35, 1.75]$ \cite{arcface}.
We can safely apply Eq. \ref{angle_function} and \ref{margin_softmax} to optimize the network when $m_1 \approx 1$ and $m_2 < 1$.

\begin{figure}[t]
  \centering
  \includegraphics[width=0.8\linewidth]{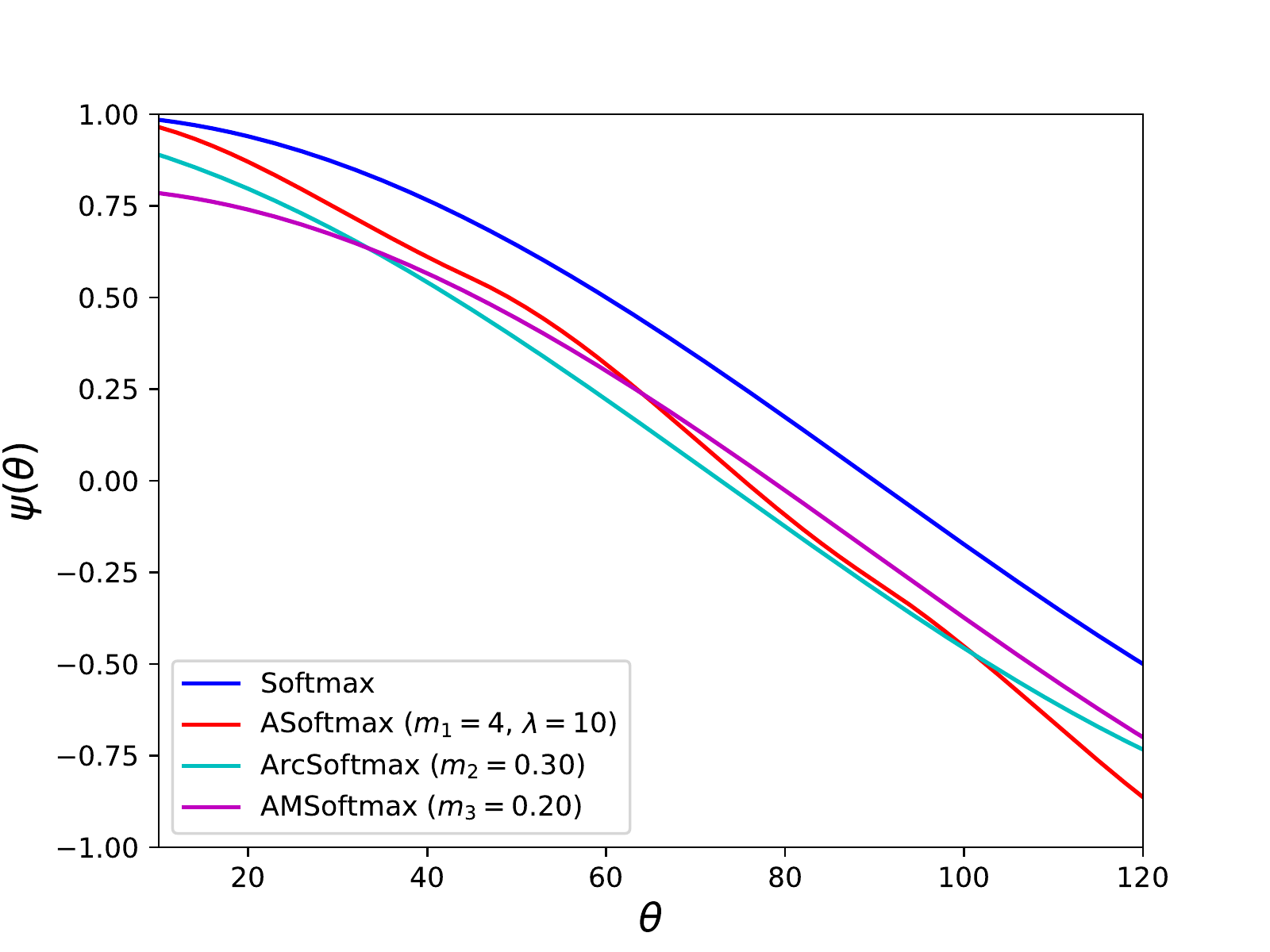}
  \caption{The angle function $\psi (\theta)$ using different margins. The configurations are selected as they performs the best in our experiments.}
  \label{fig:target_logit}
\end{figure}

When $m_1$ is large, a new $\psi(\theta_{y_i})$ is required. Let $m_2=0, m_3=0$. For $m_1 > 1.5$, we use the angle function defined in \cite{sphereface}
\begin{equation}
  \psi(\theta_{y_i}) = (-1)^k\cos (m_1 \theta_{y_i}) - 2k
  \label{asoftmax_angle_function}
\end{equation}
where $\theta_{y_i} \in [\frac{k\pi}{m_1}, \frac{(k+1)\pi}{m_1}]$ and $k\in [0, m_1-1]$.
The curves of the angle functions using different margins are illustrated in Fig. \ref{fig:target_logit}.

The margins $m_1$, $m_2$ and $m_3$ can be used separately \cite{sphereface, cosface, amsoftmax, arcface}.
Also, they can be further combined as in \cite{arcface}.
In this paper, we only use one single margin at a time since the performance gain of the margin combination is relatively small while large efforts will be paid to tune the hyperparameters.
When $m_1$, $m_2$ and $m_3$ are used individually, the losses are denoted as angular softmax (\textit{ASoftmax}), additive angular margin softmax (\textit{ArcSoftmax}) and additive margin softmax loss (\textit{AMSoftmax}), respectively.

\subsection{Feature normalization}
As discussed in \cite{l2_norm}, the $L_2$ norm of the feature is related with the sample quality when the softmax loss is used.
The network will minimize the loss by simply increasing the norm of the features for easy samples and ignoring the hard ones. This avoids the network processing samples in poor quality well.

To solve this issue, feature normalization is presented in many works. After normalization, the feature norm is eliminated from the loss and a fixed-value scaling factor $s$ is used instead.
Using the feature normalization, the loss is only related with the angle function.
Features with small norm will get much bigger gradients compared to those with large norm, making the network pay more attention to the low-quality samples \cite{amsoftmax}.

Rather than learning to map the samples into a fixed-norm hypersphere, feature normalization uses an additional normalization layer to do this job.
Unlike feature normalization, we introduce Ring loss \cite{ring_loss} to directly apply the norm constraint on the features in this paper.
The definition of the Ring loss is straightforward. We want the feature norm to be close to a target value $R$. An auxiliary loss is employed as
\begin{equation}
  L_R = \frac{\lambda_R}{N} \sum_{i=1}^{N} (\lVert \vec{x}_i \rVert - R)^2
  \label{ring_loss}
\end{equation}
where $\lambda_R$ is the loss weight with the primary large margin softmax loss. The Ring loss can be considered as a \textit{soft} version of feature normalization and the target norm $R$ can be learnt during the network training.

\subsection{Enlarge inter-speaker feature separability}
Although the large margin softmax loss improves the intra-class compactness, it does not explicitly promote the inter-class separability.
In \cite{mhe}, the authors proposed a minimum hyperspherical energy (MHE) criterion to encourage the weights of the output layer to distribute evenly on hypersphere.
The MHE criterion is expressed as
\begin{equation}
  L_M = \frac{\lambda_M}{N(C-1)} \sum_{i=1}^{N} \sum_{j=1,j \neq i}^{C} f(\lVert \hat{\vec{w}}_{y_i} - \hat{\vec{w}}_j \rVert)
  \label{ring_loss}
\end{equation}
where $\lambda_M$ is a weighting hyperparameter, $\hat{\vec{w}}_{y_i}$, $\hat{\vec{w}}_j$ are the $L_2$ normalized weights in the loss function and $f(z) = 1/z^2$ is a decreasing function. Intuitively, MHE loss enlarges the overall inter-class feature separability. Similar with the Ring loss, we include MHE as an auxiliary objective function.

\subsection{Annealing strategy during training}
From a classification perspective, the large margin softmax makes the decision boundary more stringent to correctly classify $\vec{x}_i$.
The angle between $\vec{x}_i$ and $\vec{w}_{y_i}$ is required to be much smaller than the angles with other weights.
From a view of optimization, the existence of the margin makes the well-separated features continue to get big gradients which can shrink the intra-class variance.

However, the margin will also increase the training difficulty especially when the network is randomly initialized.
To stable the training procedure, an annealing strategy is applied.
The target logit is replaced by the weighted average of the original logit and the large margin counterpart, which means
\begin{equation}
  \psi_{\text{train}}(\theta_{y_i}) = \frac{1}{1+\lambda} \psi(\theta_{y_i}) + \frac{\lambda}{1+\lambda} \cos (\theta_{y_i})
  \label{logit_average}
\end{equation}
where $\lambda=\max \{\lambda_0, \lambda_b \cdot (1+\gamma \cdot s) ^ {-\alpha}\}$, $s$ is the training step, $\lambda_0$ is the minimum value it can achieve, $\lambda_b$, $\gamma$ and $\alpha$ are the hyperparameters controlling the annealing speed.

\subsection{Other discussions}
In \cite{ge2e}, generalized end-to-end (GE2E) loss is proposed to train the speaker network. We rewrite Eq. 6 in \cite{ge2e} as
\begin{equation}
  L_{GE2E} = - \frac{1}{N} \sum_{i=1}^{N} \log \frac{e^{s \cdot \cos (\vec{x}_i, \vec{c}_{y_i}) + b}}{ \sum_{j=1}^{C'} e^{s \cdot \cos (\vec{x}_i, \vec{c}_j) + b}}
  \label{ge2e}
\end{equation}
where $C'$ is the number of speakers in a minibatch and $\vec{c}_j$ is the center of speaker $j$ estimated from the batch. 
The softmax is computed across the batch rather than the entire dataset. 
This is convenient when the training set is extremely large. 
If the bias $b$ is omitted and the estimated center $\vec{c}_j$ is replaced with a learnable weight $\vec{w}_j$, the GE2E loss becomes the modified softmax in Eq. \ref{modified_softmax}. Hence, combining with the GE2E loss, the large margin softmax loss is also potential to be applied on a dataset comprising millions of speakers.

\section{Experimental setup}
\subsection{Dataset}
To investigate the performance of the large margin softmax loss, we have run experiments on the VoxCeleb dataset \cite{voxceleb, voxceleb2}. The training set includes VoxCeleb1 \textit{dev} part and VoxCeleb2.
The VoxCeleb1 \textit{test} part is used as the evaluation set.
This setup is selected to be consistent with the Kaldi recipe \cite{kaldi}.

Equal error rate (EER), minimum detection cost function from NIST SRE08 (minDCF08) \cite{Martin_sre08} and SRE10 (minDCF10) \cite{Martin_sre10} are presented to demonstrate the performance.

\subsection{Training details}
The acoustic feature in our experiments is 30-dim MFCCs with cepstral mean normalization. 
An energy-based voice active detection (VAD) is applied. 
The training data is augmented using MUSAN \cite{musan} and RIR \cite{rir}.

We use the same network architecture as Kaldi \cite{xvector} to extract x-vectors with the following modifications.
\begin{itemize}[leftmargin=*]
\item For the frame-level network, a 5-layer TDNN is used. The kernel size for each layer is $[5,5,7,1,1]$. Different with Kaldi, there is no dilation used. This performs better in our experiments and is also suggested in other works \cite{but_system}. Statistics pooling and a 2-layer segment-level network is appended after the frame-level network.
\item Each hidden layer consists of an affine component followed by batch-normalization (BN) and ReLU non-linearity. The order of BN and ReLU does not necessarily lead to better performance but the training is more stable than that of the opposite order.
\item The last ReLU in the segment-level network is removed. The non-linearity limits the feasible angles between the feature and the weights which is not a good choice when the angle-based large margin softmax loss is applied \cite{sphereface}.
\end{itemize}

At every training step, we sample 64 speakers. For each speaker, a segment with 200 to 400 frames is sliced from the utterances.
Softmax with cross entropy is used to train the baseline system. $L_2$ regularization is applied to all layers in the network to prevent overfitting. We select stochastic gradient descent (SGD) as the optimizer and the initial learning rate is set to 0.01. A 1000-utterance validation set is randomly selected from the training set and the learning rate is halved if the validation loss gets stuck for a while. The loss converges after the learning rate goes down below $10^{-5}$, resulting to around 2.5M training steps. No dropout is applied in our networks as described in \cite{but_system}.

When training with the large margin softmax loss, a $\lambda$ annealing strategy is used as described in Section 3.4. Specifically, we set a fast decay for AMSoftmax where $\lambda_b=1000, \gamma=0.0001, \alpha=5$. For ASoftmax and ArcSoftmax, the decay speed is slowed down by setting $\gamma=0.00001$. The $\lambda_0$ for ArcSoftmax and AMSoftmax is 0, while for ASoftmax, $\lambda_0 = 10$. This non-zero $\lambda$ results in a more gentle angle function. For example, when $m_1=4, \lambda=10$, the target logit is similar to ASoftmax with $m_1=1.2$ without $\lambda$.

After training, the output of the second last layer in the segment-level network is extracted as the speaker embedding. LDA is used to reduce the dimension to 200 and PLDA is then applied to generate the verification scores. One may also use the embedding extracted from the last layer with simple cosine backend. However, in our experiments, we find the PLDA scoring generally performs better.

Our systems are implemented with Kaldi and Tensorflow toolkits. The code and models have been released \footnote{https://github.com/mycrazycracy/tf-kaldi-speaker}.

\section{Results}
Table \ref{tab:result} summarizes the results of different systems.
The first row in Table \ref{tab:result} shows the Kaldi recipe for VoxCeleb.
The first experiment is to validate the performance of our baseline system. 
We find a large weight decay parameter works well in our systems. 
When increasing this parameter from $0.0001$ to $0.01$, the EER is improved from 3\% to 2.34\%. 
The second row shows the performance of our baseline system which is significantly outperform the standard Kaldi result.
The third row is the result using the modified softmax loss. 
Without any margins, the modified softmax does not perform better by simply normalizing the weights.

The performance of the large margin softmax loss is exhibited in the following sections of Table \ref{tab:result}. 
We remove the last ReLU for all these networks. This generally improves the results. 
For instance, the performance of ASoftmax ($m_1$=4) with ReLU is 2.12\%, 0.0122 and 0.3214 in EER, minDCF08 and minDCF10, while without ReLU, it achieves 2.15\%, 0.0113 and 0.3108 instead. The same trends are observed in other systems as well.

From Table \ref{tab:result}, it is clear that ASoftmax achieves the best result when $m_1=4$.
The performance of ArcSoftmax is similar with ASoftmax and the best margin is about $m_2=0.25$ to $m_2=0.30$.
The AMSoftmax performs the best among all these large margin softmax losses with the optimal margin $m_3=0.20$.
We notice that the best margins for these systems are relatively small compared with those reported in the face verification \cite{sphereface, amsoftmax, arcface}.

\begin{figure}[t]
  \centering
  \includegraphics[width=0.7\linewidth]{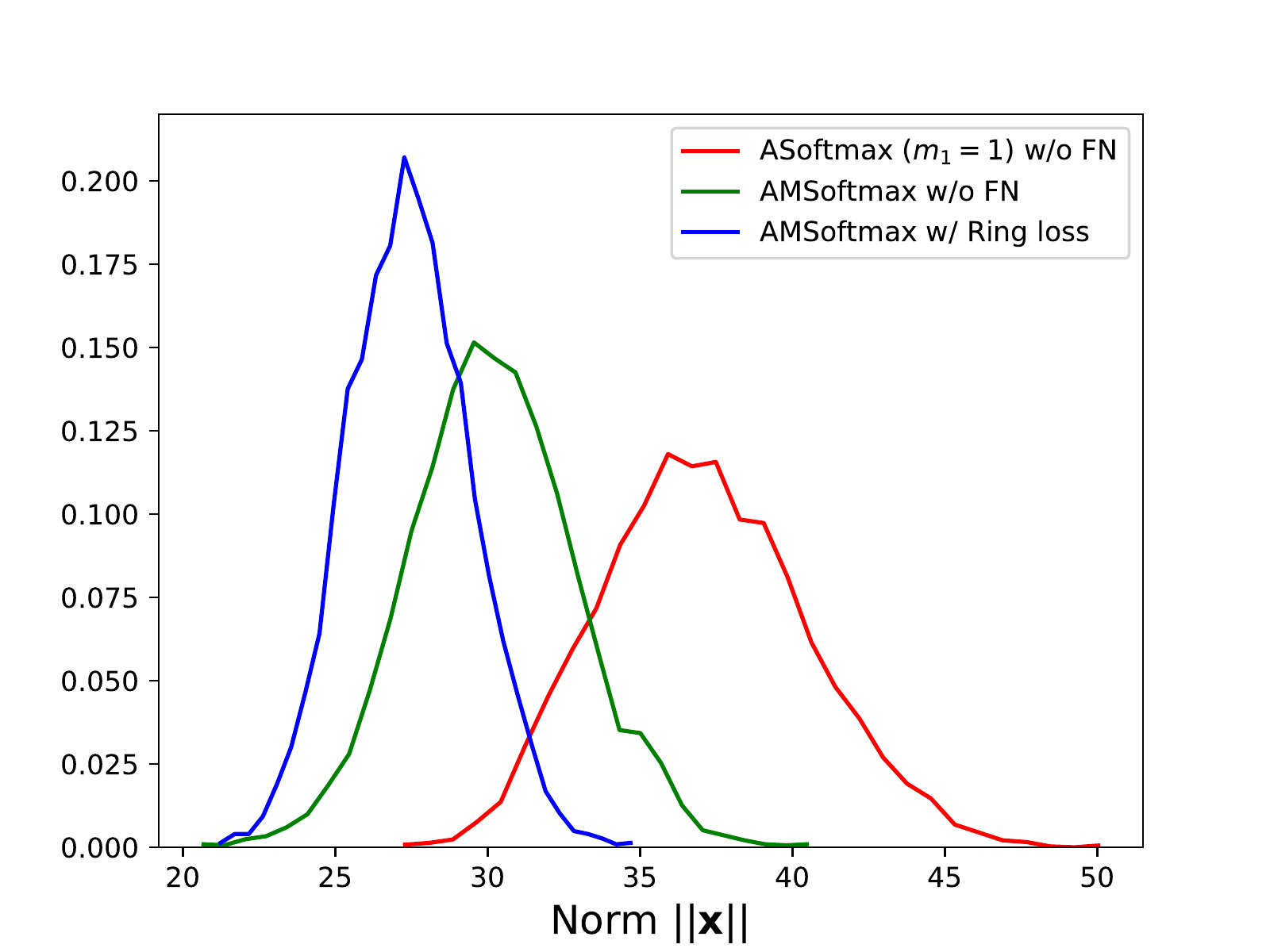}
  \caption{The norm distribution on the test set. When the Ring loss is used, the norm variance becomes smaller.}
  \label{fig:norm_dist}
\end{figure}

\begin{figure}[t]
  \centering
  \includegraphics[width=0.7\linewidth]{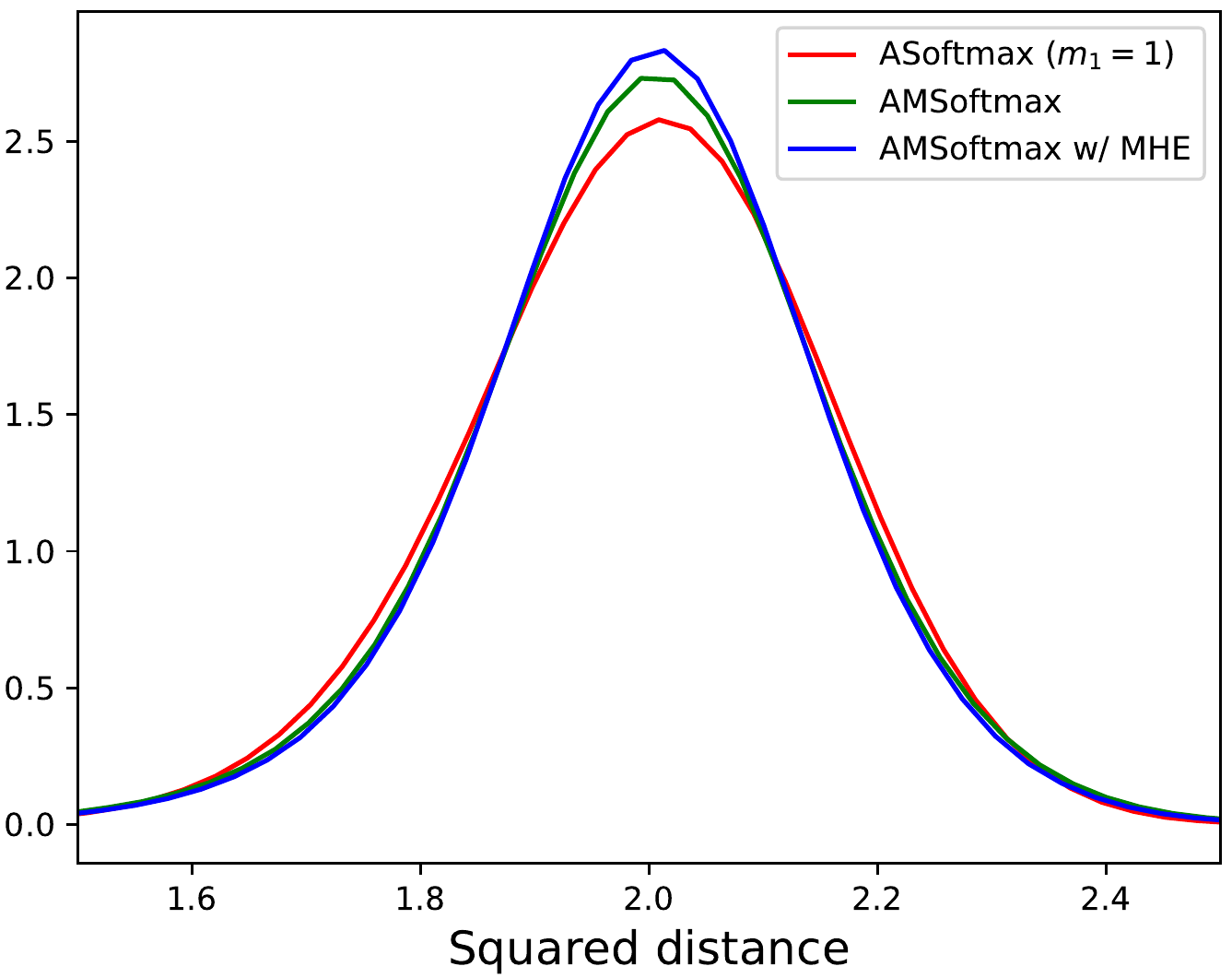}
  \caption{The distribution of the squared distances between the normalized weights in the output layer. The means for all the systems is about 2.0 and the MHE loss reduces the variance.}

  \label{fig:inter_class_dist}
\end{figure}

We now investigate the influence of the Ring loss and the MHE loss.
The weight $\lambda_R$ for the Ring loss is set as 0.01 and $R$ is initialized at 20.
Table \ref{tab:result} shows that the Ring loss improves the minDCF08 and minDCF10.
The norm distributions of different systems are presented in Fig. \ref{fig:norm_dist}.
Since the weights of the softmax network are not normalized, the mean of the feature norm is very large (about 150). 
To show the feature norm without margins, we use the modified softmax instead. 
From Fig. \ref{fig:norm_dist}, we find that using the margin helps to reduce the norm variance.
The margin prevents the norm of the simple samples growing too large.
The norm distribution further shrinks when the Ring loss is applied.
However, even though the network is trained using AMSoftmax without feature normalization, the norm variance is relatively small. Therefore, the effectiveness of the Ring loss with AMSoftmax is less significant in our experiment.

The performance of the MHE loss is presented in the last row of Table \ref{tab:result}.
The weight $\lambda_{M}$ is 0.01.
The AMSoftmax with MHE loss achieves the best result among all the systems.
This loss improves the baseline performance by 15\% in EER, 13\% in minDCF08 and 33\% in minDCF10. 
To get some insights of the MHE loss, we illustrate the distribution of the pairwise squared distances between the normalized weights. 
The distance, which is $\lVert \hat{\vec{w}}_i - \hat{\vec{w}}_j\rVert^2, \forall i,j, i\neq j$, indicates the separability between speakers on the training set.
In Fig. \ref{fig:inter_class_dist}, it is shown that all the distributions have the means at about 2.0, indicating $\cos (\hat{\vec{w}}_i, \hat{\vec{w}}_j)$ is 0 in average.
The AMSoftmax with the MHE loss achieves the smallest variance of the inter-speaker distances, which means the features of speakers distribute more evenly on hypersphere, leading to a better overall separability in the feature space.

\begin{table}[t]
  \caption{The comparison of results between systems training with different loss functions. The weights for the Ring loss and the MHE loss are both 0.01.}
  \label{tab:result}
  \centering
  \begin{tabular}{lcccc}
  \toprule
  \multicolumn{2}{c}{} & EER(\%) & minDCF08 & minDCF10 \\
  \midrule
  \multicolumn{2}{l}{Kaldi \footnote{https://github.com/kaldi-asr/kaldi/blob/master/egs/voxceleb/v2}} & 3.10 & 0.0169 & 0.4977 \\
  \multicolumn{2}{l}{Softmax} & 2.34 & 0.0122 & 0.3754 \\
  \multicolumn{2}{l}{Modified Softmax} & 2.62& 0.0131 & 0.4146 \\
  \midrule
  \multirow{2}{*}{ASoftmax} & $m_1=2$ & 2.18 & 0.0119 & 0.3791  \\
  & $m_1=4$ & 2.15 & 0.0113 & 0.3108 \\
  \midrule
  \multirow{4}{*}{ArcSoftmax} & $m_2=0.20$ & 2.14 & 0.0119 & 0.3610 \\
  & $m_2=0.25$ & 2.03 & 0.0120 & 0.4010 \\
  & $m_2=0.30$ & 2.12 & 0.0115 & 0.3138 \\
  & $m_2=0.35$ & 2.23 & 0.0123 & 0.3622 \\
  \midrule
  \multirow{4}{*}{AMSoftmax} & $m_3=0.15$ & 2.13 & 0.0113 & 0.3707 \\
  & $m_3=0.20$ & 2.04 & 0.0111 & 0.2922 \\
  & $m_3=0.25$ & 2.15 & 0.0119 & 0.3559 \\
  & $m_3=0.30$ & 2.18 & 0.0115 & 0.3152 \\
  \midrule
  \multicolumn{2}{l}{AMSoftmax + Ring Loss} & 2.07 & 0.0107 & 0.2687 \\
  \multicolumn{2}{l}{AMSoftmax + MHE} & \textbf{2.00} & \textbf{0.0106} & \textbf{0.2487} \\
  \bottomrule
  \end{tabular}
\end{table}

\section{Conclusions}
In this paper, we investigate the large margin softmax loss for speaker verification. 
By selecting an appropriate margin, the large margin softmax loss can achieve promising results.
Ring loss and MHE loss are involved to further improve the performance.
Ring loss is a soft version of feature normalization and alleviates the impact of feature norm. 
MHE criterion is another loss function which enlarges the overall inter-speaker separability.
On VoxCeleb, our baseline system achieves better result than the Kaldi toolkit. 
We find AMSoftmax is easier to train and generally performs better than ASoftmax and ArcSoftmax in our experiments. 
The best system is obtained when AMSoftmax is used with the MHE loss. 
This combination substantially outperforms the baseline.

In the future, we will combine both the Ring loss and the MHE loss with the large margin softmax loss. More efforts will be made to enable simple cosine scoring and remove the need for the PLDA backend.

\section{Acknowledgements}
This work was supported by the National Natural Science Foundation of China under Grant No. 61403224 and No. U1836219.

\bibliographystyle{IEEEtran}
\bibliography{mybib}

\end{document}